\documentclass[twoside, epsfig]{article}

\usepackage[round]{natbib}

\usepackage{graphicx}
\usepackage{color}
\usepackage{array}
\usepackage{lscape}
\usepackage{rotating}
\usepackage{longtable}
\usepackage{multirow}
\usepackage{fancyhdr}
\usepackage[colorlinks=true,urlcolor=blue,citecolor=black,bookmarks=false]{hyperref}

\font\ftitle=cmssbx14
\font\fabs=cmr10

\usepackage{amssymb}
\usepackage{makecell}
\usepackage{enumitem}

\def\MyTitleTwo[#1]#2 {
  {\centerline {\ftitle #1} \vskip 0.3cm
   \centerline {\ftitle #2} \vskip 0.5cm plus 0.4cm minus 0.2cm}
}

\def\MyAuth#1 {
 \centerline {#1}
 \vskip 0.5cm
}

\def\MyInst#1{ 
 {\centering \small #1 \vskip 1mm }\large
}

\def\begintext{
    \vskip 0.5cm
    \baselineskip=5mm
    \noindent
}

\def\MyAbstract#1{
    \vskip 0.5cm
    \begin{center}
    \begin{tabular}{p{0.9\textwidth}}
    {\fabs {\bf Abstract:} #1 }\\
    \end{tabular}
    \end{center}
}

\def\MyAcknowledgements#1{
    \vskip 0.5cm
    \noindent{\fabs{\bf Acknowledgements:} #1 }\\
}

\def\MyBibcode#1{
    \href{https://ui.adsabs.harvard.edu/abs/#1}{\underline{#1}}
}

\def\MyLink#1{
     \href{#1}{\underline{#1}}
}

\begin{document}

\MyTitleTwo[Six New Variable Stars Discovered from Ground-Based]{Photometry and Characterized with TESS Data}
\MyAuth{Maksym Yu. Pyatnytskyy}
\MyInst{ Private Observatory ”Osokorky” Kyiv, Ukraine, {\tt \href{mailto:pmak@osokorky-observatory.com}{pmak@osokorky-observatory.com}}}

\MyAbstract{
We report the discovery of six new variable stars identified through an exploratory analysis of several sky fields observed by the author using a small telescope and a CMOS camera. The search employed simultaneous photometry of hundreds of objects with AstroImageJ, supplemented by custom Python-based tools developed by the author to generate aperture lists and visualize light curves across the full field of view. The variables were further characterized using TESS data, and their properties are presented. We also describe the data analysis workflow, including the software packages LCV and VS-fit, developed by the author for periodogram analysis and light-curve modelling.
}

\begintext

\section{Introduction}\label{secintro}

The discovery of new variable stars has become increasingly difficult in the era of automated sky surveys. For example, the Gaia DR3 catalog \citep{gaia2022b} contains about 10 million classified variable sources, the majority of which were identified for the first time by Gaia. The aim of this work is to assess the capability of a small ground-based telescope in an urban environment to detect variable stars, and to demonstrate the use of TESS data for their detailed characterization. The detection is based on our own ground-based photometric observations, while the characterization relies primarily on TESS data.

\section{Observation and data reduction}

The observation site was located in the Kyiv urban area, in the Osokorky neighborhood, on the left bank of the Dnipro River. According to the light pollution map, the observing site corresponds to Bortle class 7.

All observations were conducted using a small Newtonian telescope with an aperture of 150 mm and a focal length of 750 mm. The telescope was equipped with a ZWO ASI183MM Pro monochrome cooled CMOS camera and a filter wheel. It was mounted on an equatorial mount with guiding.

The light frames in FITS format were processed using a standard calibration procedure, including dark, flat, and dark-flat frames. The author used his own command-line toolkit to calibrate the images: \MyLink{https://github.com/mpyat2/MaxFITStoolkit}. The calibrated light frames were then binned in software by a factor of 2, reducing their size and facilitating further processing. The calibration frames were acquired each night immediately after the observing session. Flat-field frames were obtained using a white paper screen placed approximately 2 m from the telescope and illuminated by an LED lamp.

The field of view of the setup was approximately 40 × 60 arcmin. Three different star fields were observed; they are summarized in Tab.~\ref{tableFields}. In the table, the TESS Input Catalog identifiers (TIC) for the newly discovered stars are given, along with their primary names in the International Variable Star Index (VSX) \citep{watson2006}.

\begin{table}
\caption{Summary of observed star fields.}\vspace{3mm}
\centering
\begin{tabular}{ccccc}
\hline
	 \makecell{Field center \\ (J2000) {[} h:m +$^{\circ}$:$'${]}} & Filter & \makecell{Session \\ duration \\ {[}hours{]}} & \makecell{Individual \\ exposure \\ {[}s{]}} & New variables found \\ \hline \hline
  23:26 +42:32 & clear & 3.7 & 60 & \makecell[lt]{TIC 174042954 = PMAK V147\\[1mm]} \\
  06:01 +27:48 & Johnson V & 3.8 & 30 & \makecell[lt]{TIC 79234006 = PMAK V148 \\ TIC 79234043 = PMAK V149\\[1mm]} \\
  23:21 +61:12 & clear & 3.5 & 60 & \makecell[lt]{TIC 269112631 = PMAK V150 \\ TIC 269809093 = PMAK V151 \\ TIC 269112450 = PMAK V152} \\
\hline
\end{tabular}\label{tableFields}
\end{table}

\section{Data analysis}\label{secdata}

\subsection{Building light curves from our observations}

Photometry of the FITS images obtained for each observed star field was performed using the AstroImageJ software \citep{collins2017}. In addition, two Python scripts developed by the author were used as auxiliary tools. The scripts, together with documentation, are available at \MyLink{https://github.com/mpyat2/MaxFITStoolkit/tree/master/PyScripts}.

The first script, apgen.py, was used to extract star-like sources from a FITS image--one of the images in a set of frames of the star field to be measured. The script utilizes the DAOStarFinder algorithm \citep{stetson1987}, as implemented in the Photutils package \citep{astropy2013, astropy2018, astropy2022, photutils2024}. It outputs a file with the coordinates of the sources. At least one of the sources must be chosen as a comparison star with a known magnitude. A set of calibrated FITS images of the star field was loaded into AstroImageJ, and a list of objects to be measured was then loaded from the coordinate file. The practical limit for the number of apertures measured simultaneously is about 1000.

AstroImageJ produces an output table that contains magnitude information for each source in all frames.

The second script, aijplot.py, was used to visualize light curves. It takes a table with magnitude information produced by AstroImageJ and outputs an HTML file containing light-curve plots. Data binning can be applied by averaging every N points to reduce noise; N can be specified as a command-line parameter. The resulting HTML file can then be loaded into a browser, where the light curves can be visually inspected.

\begin{figure}[htbp]
\centering
\includegraphics[width=10cm]{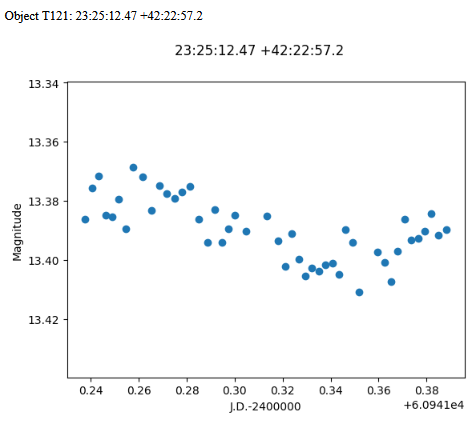}
\caption{An example light curve derived from our observations. The target is TIC 174042954, listed in the VSX catalog as PMAK V147.}
\label{figLCexample}
\end{figure}

Through this visual inspection, many known short-period variables were detected. Among them, however, some light curves appeared to belong to previously unregistered variables. An example of such a light curve is shown in Fig.~\ref{figLCexample}. In this example, the data were binned with a bin size of 4 points.

This method allows one to find short-period variables or variables experiencing rapid changes. However, identification of the variability type, refinement of parameters, etc., require additional data. Currently, a rich source of variability data is the TESS survey \citep{ricker2014}.

% png format

\subsection{Retrieving TESS light curves}

TESS data are available in different forms, including pre-calculated light curves produced by various pipelines, such as the SPOC pipeline (Science Processing Operations Center) \citep{jenkins2016}, the MIT Quick-Look Pipeline (QLP) \citep{huang2020}, and others.

We used the Lightkurve Python package \citep{lightkurve2018} to download the TESS light curves. To simplify the process, a text-mode, menu-driven utility, TESSdata.py, was developed. It is available at \\ \MyLink{https://github.com/mpyat2/LightCurveViewer/tree/main/scripts/tess\_data/}, along with the documentation.

Pre-calculated TESS light curves are not available for all stars, as SPOC products are restricted to pre-selected targets, while Full Frame Image (FFI)-based pipelines such as QLP become progressively less complete toward fainter magnitudes. In cases where pre-calculated light curves are not available, we used Lightkurve to extract light curves from TESS Full Frame Images (FFIs) using aperture photometry.

\subsection{Analysis of light curves}

All variables detected in this survey were periodic, which implies the need for periodogram analysis. An excellent program for performing such analysis is the MCV software \citep{andronov2004}. For periodogram analysis, it can use not only simple harmonic functions but also their harmonics, as described in \citep{andronov1994, andronov2020}. The software can also approximate light curves using combinations of algebraic and trigonometric polynomials of different degrees, with estimation of parameter uncertainties, including period refinement and associated error estimates.

However, the MCV program was developed quite a while ago and does not utilize modern computational features such as multiprocessing, which can significantly improve the speed of calculations. This is why we developed a new program, Light Curve Viewer (LCV).

The LCV program is open source and available on GitHub: \\
\MyLink{https://github.com/mpyat2/LightCurveViewer}.
The program has limited cross-platform support: it is compiled for Windows and Debian-based Linux distributions. For optimal performance in computing approximations using trigonometric and algebraic polynomials, which are also used in periodogram analysis, the Intel oneMKL library is used \citep{intel_onemkl}.

Key features of the LCV program are:
\begin{itemize}[nosep]
\item Visualization of light curves from plain text files
\item Building phase plots
\item Calculation of periodograms using trigonometric polynomials of different degrees, with optional combination with an algebraic polynomial.
 
This means that for each sample frequency $f$, the following function is used to approximate the light curve:

\begin{equation}\label{eqPoly}
y_c(t) = \sum_{\alpha=0}^{r} C_{1\alpha} t^{\alpha} + \sum_{\beta=1}^{s} \left[ C_{2\beta} \sin (\beta \omega t) + C_{3\beta} \cos(\beta \omega t) \right]
\end{equation}

Here, $\omega = 2\pi f$, $r$ is the degree of the algebraic polynomial (reducing to a constant for $r=0$), and $s$ is the degree of the trigonometric polynomial.

We approximate the observations $y_k$, obtained at times $t_k$, by the function $y_c(t)$ using the least-squares method; the coefficients $C_{1\alpha}$, $C_{2\beta}$, and $C_{3\beta}$ are the parameters of the approximation.

The following statistic is then calculated:
\begin{equation}\label{eqStatistics}
S(f) = 1 - \frac{\sigma_{O-C}^2}{\sigma_O^2}
\end{equation}

where $\sigma_O^2$ is the r.m.s. deviation of the observations $y_k$ (observed magnitudes) from the mean value\footnote{If the degree of the algebraic polynomial is zero; if the degree is greater than zero, a reference polynomial approximation is calculated for the observations.}, and $\sigma_{O-C}^2$ is the r.m.s. deviation of the observations $y_k$ from the calculated values $y_c(t_k)$ (see \citep{andronov1994} for details).
This statistic is referred to as \textit{power} in LCV.

The degrees of both the trigonometric and algebraic polynomials are specified as parameters in the periodogram calculation.

Note that in its simplest form, where $r = 0$ and $s = 1$, the periodogram calculated in this way is equivalent to the Ferraz-Mello DC DFT \citep{ferrazmello1981} (\textit{power} in the Ferraz-Mello periodogram is the statistic $S(f)$ multiplied by the factor $(n - 1)/2$, where $n$ is the number of observations).

During these calculations, the program can utilize part or all of the available processor cores.
\item Approximation of light curves using combinations of algebraic and trigonometric polynomials of varying degrees and frequencies (periods).
\item Detrending of light curves
\end{itemize}

Unlike MCV, the LCV program cannot refine periods while approximating using trigonometric polynomials. To fill this gap, another Python utility with a graphical user interface was developed: VS-fit. It is available on GitHub: \MyLink{https://github.com/mpyat2/VS-fit}. The utility uses the scipy.optimize package \citep{virtanen2020scipy} and can refine periods for a combination of up to nine trigonometric polynomials of different degrees, with an optional algebraic polynomial. Estimation of the associated errors is also performed.

\section{Results}

Six new periodic variables were discovered as a result of visual inspection of light curves derived from observations of star fields listed in Tab.~\ref{tableFields}.

Tab.~\ref{tableStars} summarizes the information for these variables. A value in parentheses denotes the amplitude; in such cases, the first value represents the mean magnitude. Otherwise, the values correspond to the magnitudes at maximum and minimum light. We used V magnitudes derived from Gaia DR3 data \citep{gaia2022a} as reference magnitudes.

\begin{table}
\caption{Parameters of the variable stars.}\vspace{3mm}
\centering
\begin{tabular}{cccccc}
\hline
  TIC id & \makecell{RA Dec (J2000) \\ {[}h:m:s $^{\circ}$:$'$:$''${]}} & Range & Type & \makecell{Period \\ {[}d{]}} & \makecell{Epoch  \\ {[}$BJD_{TDB}$\\$-2400000${]}}\\ \hline \hline
  174042954 & 23:25:12.41+42:22:56.1 & \makecell{13.40 V \\(0.03)TESS\\[1mm]} & ELL & 0.3150477(1) & 59870.657 \\
  79234006  & 06:00:02.75+28:09:23.5 & \makecell{11.93 V \\ (0.11)TESS\\[1mm]} & ROT & 0.424885(7) & 59492.449 \\
  79234043  & 05:59:43.57+28 08 13.9 & \makecell{12.24 V \\ (0.02)TESS\\[1mm]} & DSCT & 0.0506021(3) & 60250.1295 \\
  269112631 & 23:17:52.80+60:50:05.8 & \makecell{14.48 V \\ (0.08)TESS\\[1mm]} & EA+BCEP & \makecell{41.04236(7) EA \\ 0.13287783(9) BCEP } & \makecell{60616.2154 \\ 59898.1775}\\
  269809093 & 23:23:06.05+61:19:10.2 & \makecell{13.52 V \\ (0.07)TESS \\[1mm]} & SPB & 0.3679756(6) & 60400.827 \\
  269112450 & 23:17:46.01+60:53:19.4 & \makecell{15.24 V \\ (0.08)TESS} & EA+BY  & \makecell{1.1231069(5) EA \\ 1.117347(4) BY }    & \makecell{60422.847 \\ 60422.744}\\
\hline
\end{tabular}\label{tableStars}
\end{table}

\subsection{TIC 174042954}

\begin{figure}[htbp]
\centering
\includegraphics[width=15cm]{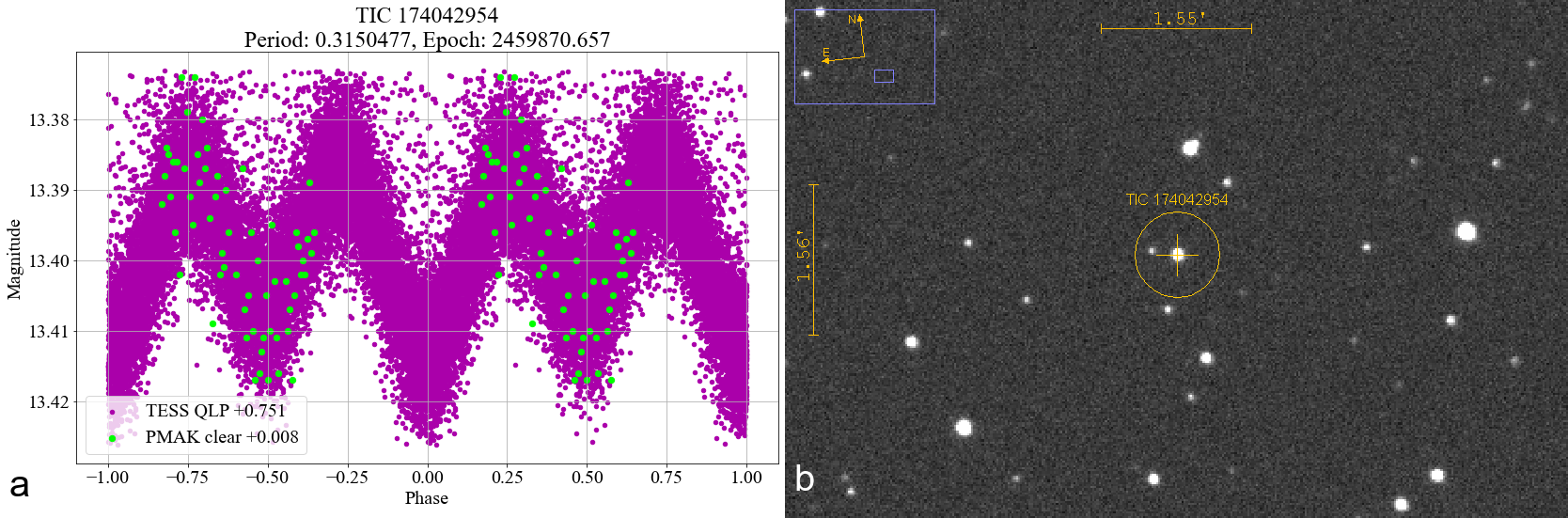}
\caption{(a) Phase-folded light curve of TIC 174042954; (b) the corresponding field.}
\label{figPMAK_V147}
\end{figure}

This star is registered in the VSX database as PMAK V147 and is classified as a rotating ellipsoidal variable (ELL). In Fig.~\ref{figPMAK_V147}a, the phase plot is shown, where our data are labeled as PMAK.
TESS QLP light curves from Sectors 57 and 84 were used, and were aligned to the median magnitude derived from Gaia DR3 data.
The period listed in Tab.~\ref{tableStars} was obtained from a periodogram analysis of the TESS data, calculated with LCV, and subsequently refined using a second-degree trigonometric polynomial with VS-fit.
In Fig.~\ref{figPMAK_V147}b, a portion of the stellar field containing this object is shown.

% png format

\subsection{TIC 79234006}

\begin{figure}[htbp]
\centering
\includegraphics[width=12cm]{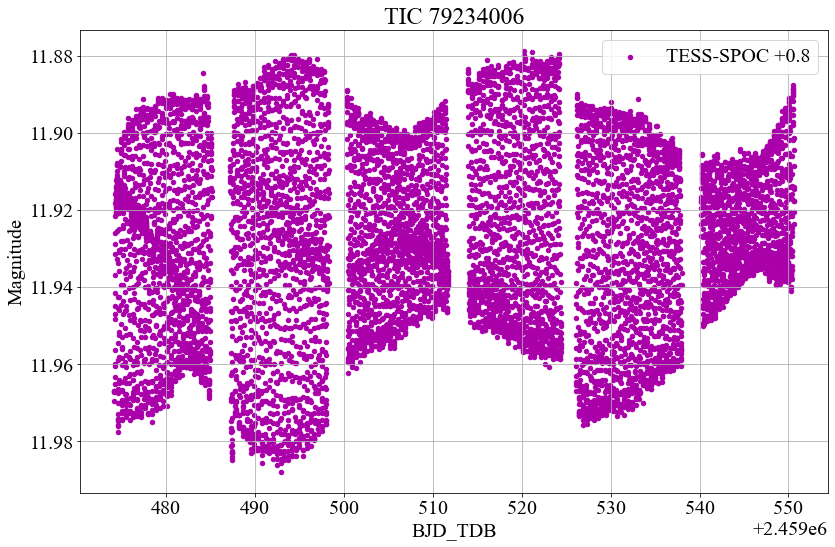}
\caption{Light curve of TIC 79234006, with the mean magnitude derived from Gaia DR3 data.}
\label{figPMAK_V148lc}
\end{figure}

This star is registered in the VSX database as PMAK V148. This is a rotational variable, as evidenced by its TESS-SPOC light curve (sectors 43, 44, 45) shown in Fig.~\ref{figPMAK_V148lc}.

\begin{figure}[htbp]
\centering
\includegraphics[width=15cm]{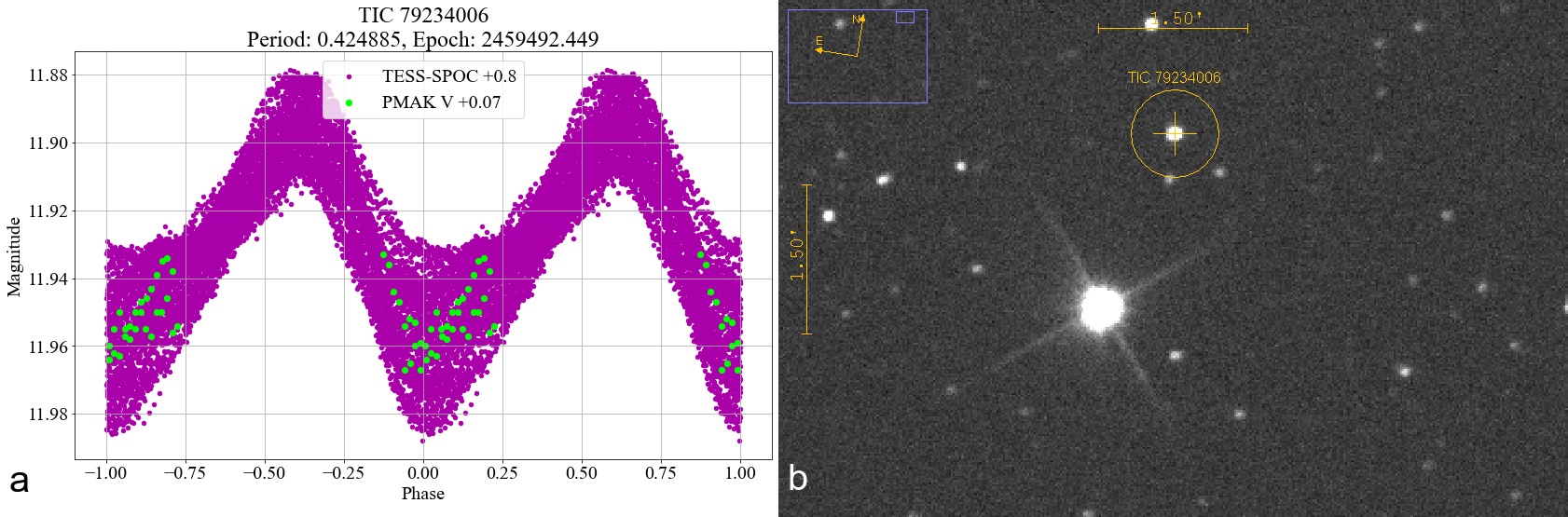}
\caption{(a) Phase-folded light curve of TIC 79234006; (b) the corresponding field.}
\label{figPMAK_V148}
\end{figure}

Periodogram analysis using LCV, followed by period refinement with VS-fit, yields the period value shown in Tab.~\ref{tableStars}. The corresponding phase plot is shown in Fig.~\ref{figPMAK_V148}a. Our observations are labeled as PMAK. Two observing sessions were conducted for this variable, on 8 and 11 March 2026. The light curves were aligned using the mean V magnitude derived from Gaia DR3 data. The corresponding star field is shown in Fig.~\ref{figPMAK_V148}b.

% png format

% png format

\subsection{TIC 79234043}

This star is registered in the VSX database as PMAK V149 and is classified as a pulsating variable of the $\delta$ Scuti type (DSCT).

\begin{figure}[htbp]
\centering
\includegraphics[width=12cm]{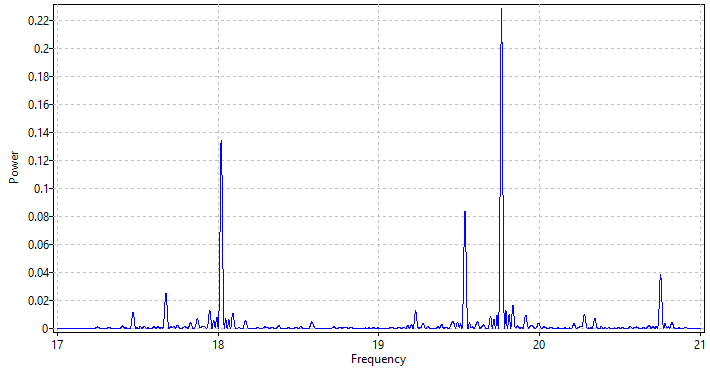}
\caption{Periodogram of TESS data (Sectors 71 and 72) for the star TIC 79234043. Frequency is in $d^{-1}$.}
\label{figPMAK_V149periodogram}
\end{figure}

The periodogram obtained with LCV from the TESS data for Sectors 71 and 72 shows several clearly visible peaks, see Fig.~\ref{figPMAK_V149periodogram}. A model was built using trigonometric polynomials with the most prominent frequencies, and the corresponding periods were refined using the VS-fit program. Tab.~\ref{tablePeriodsPMAK_V149} shows the periods and corresponding semi-amplitudes derived from the best fit.

% png format

\begin{table} % note that caption is above the table
\caption{Periods and semi-amplitudes of the best-fit model for the light curve of TIC 79234043.}\vspace{3mm}
\centering
\begin{tabular}{cc}
\hline
	 Period [d] & Semi-amplitude [mag] \\ \hline \hline
  0.0506021(3) & 0.00372(4) \\
  0.0554973(5) & 0.00300(4) \\
  0.0511909(6) & 0.00203(5) \\
  0.0481879(7) & 0.00161(4) \\
\hline
\end{tabular}\label{tablePeriodsPMAK_V149}
\end{table}

\begin{figure}[htbp]
\centering
\includegraphics[width=12cm]{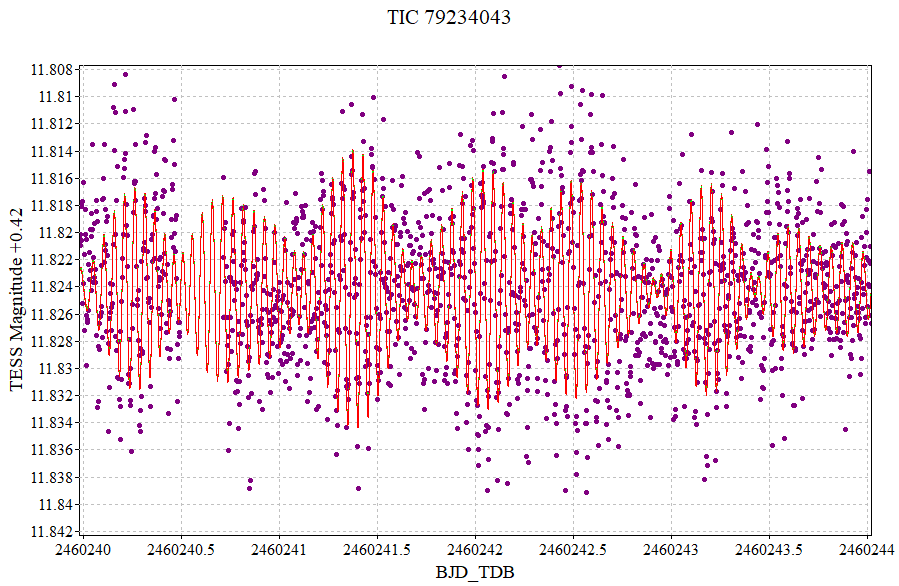}
\caption{Portion of the TIC 79234043 light curve together with the model.}
\label{figPMAK_V149lc}
\end{figure}

Fig.~\ref{figPMAK_V149lc} shows a portion of the TIC 79234043 light curve together with the light-curve model created with LCV.

% png format

\begin{figure}[htbp]
\centering
\includegraphics[width=15cm]{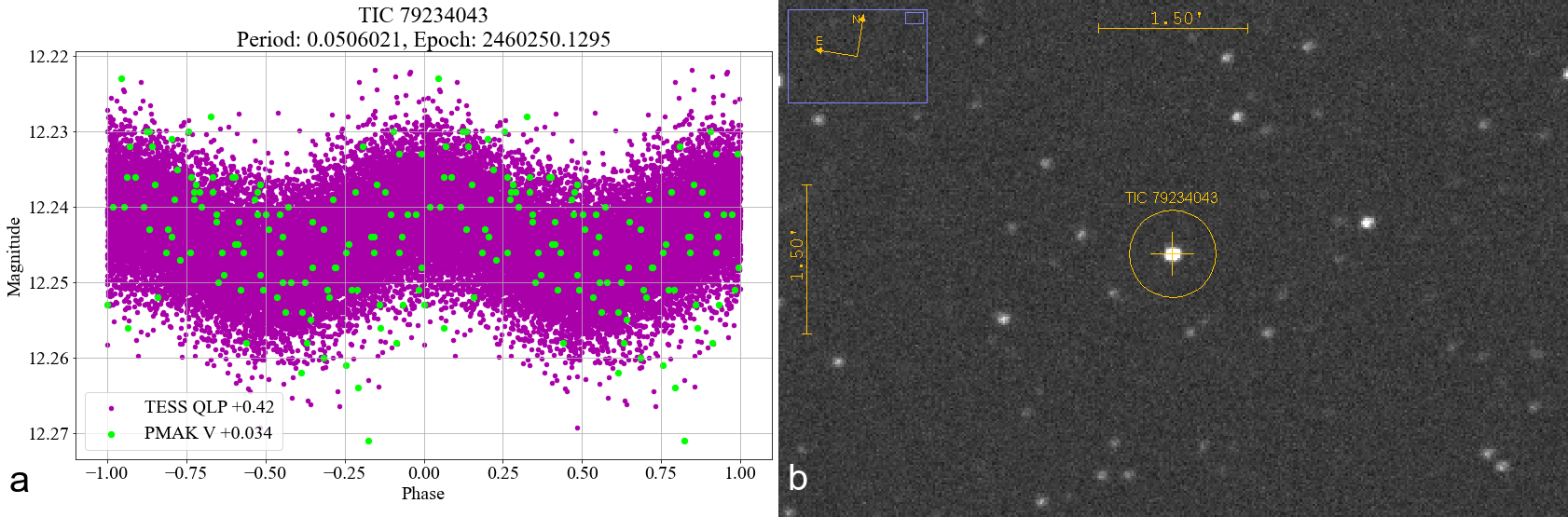}
\caption{(a) Phase-folded light curve of TIC 9234043; (b) the corresponding field.}
\label{figPMAK_V149}
\end{figure}

Phase plot built for the most prominent period of 0.0506021 d is shown in Fig.~\ref{figPMAK_V149}a. Our observations are labeled as PMAK. Three observing sessions were conducted for this variable on 8, 10, and 11 March 2026. The light curves were aligned using the mean V magnitude derived from Gaia DR3 data. The corresponding star field is shown in Fig.~\ref{figPMAK_V149}b.

% png format

\subsection{TIC 269112631}
This star is registered in the VSX database as PMAK V150. It has a combined classification: an eclipsing variable of Algol type and a pulsating variable of the $\beta$ Cephei type (EA+BCEP).

\begin{figure}[htbp]
\centering
\includegraphics[width=15cm]{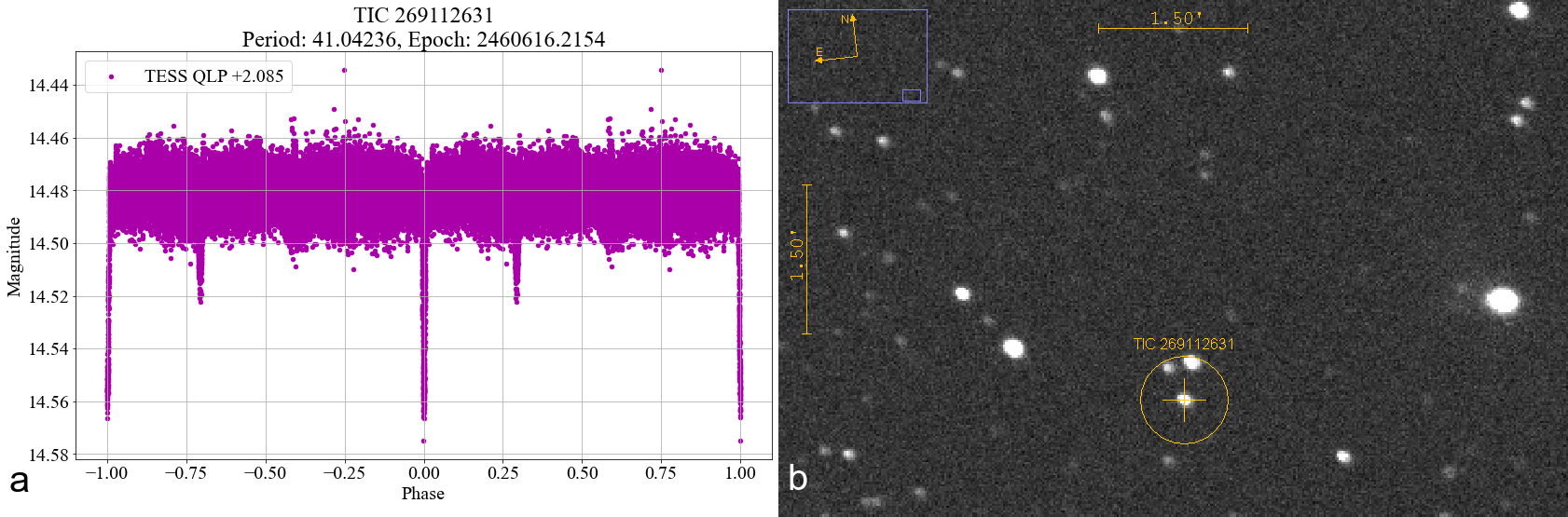}
\caption{(a) Phase-folded light curve of TIC 269112631 (EA variability); (b) the corresponding field.}
\label{figPMAK_V150}
\end{figure}

TESS observations for this star are available in Sectors 17 and 24 (effective exposure time 1426 s, cadence 1800 s) and 57, 58, 77, 78, 84, and 85 (effective exposure time 158 s, cadence 200 s). While analyzing the TESS data, we first noticed sharp eclipsing minima. The phase plot with the eclipse period of 41.04236(7) days is shown in Fig.~\ref{figPMAK_V150}a. The light curves were aligned using the mean V magnitude derived from Gaia DR3 data. The corresponding star field is shown in Fig.~\ref{figPMAK_V150}b.

% png format

\begin{figure}[htbp]
\centering
\includegraphics[width=15cm]{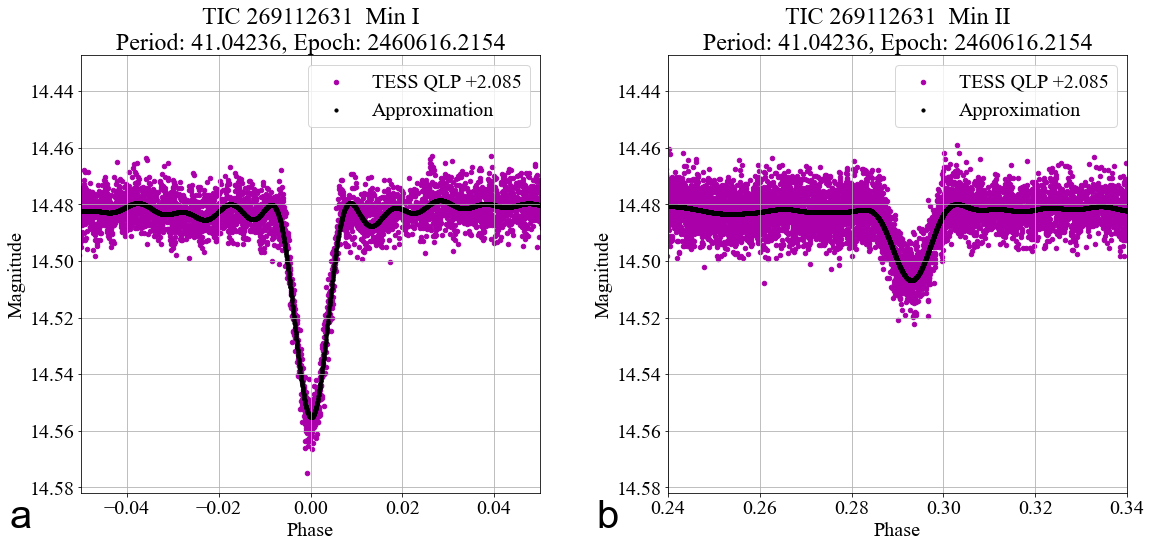}
\caption{Zoomed phase-folded light curve of TIC 269112631 near the primary (a) and secondary (b) minima.}
\label{figPMAK_V150minima}
\end{figure}

We estimated the period and its associated error using the VS-fit software. The light curve was approximated with a trigonometric polynomial of high degree, starting from 20, with subsequent refinement of the period. An initial value of the period (41.0422 d) was estimated “by eye” from the phase plot. At a degree of 100, we obtained a satisfactory approximation of both narrow minima; see Fig.~\ref{figPMAK_V150minima}. A further increase in the polynomial degree (up to 200) resulted in only small changes in the estimated period, smaller than the estimated error.

% png format

\begin{figure}[htbp]
\centering
\includegraphics[width=12cm]{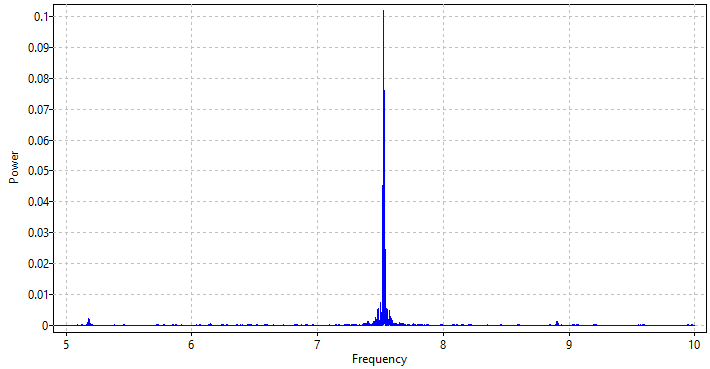}
\caption{Periodogram for TIC 269112631 excluding the eclipsing minima.}
\label{figPMAK_V150periodogram}
\end{figure}

\begin{figure}[htbp]
\centering
\includegraphics[width=12cm]{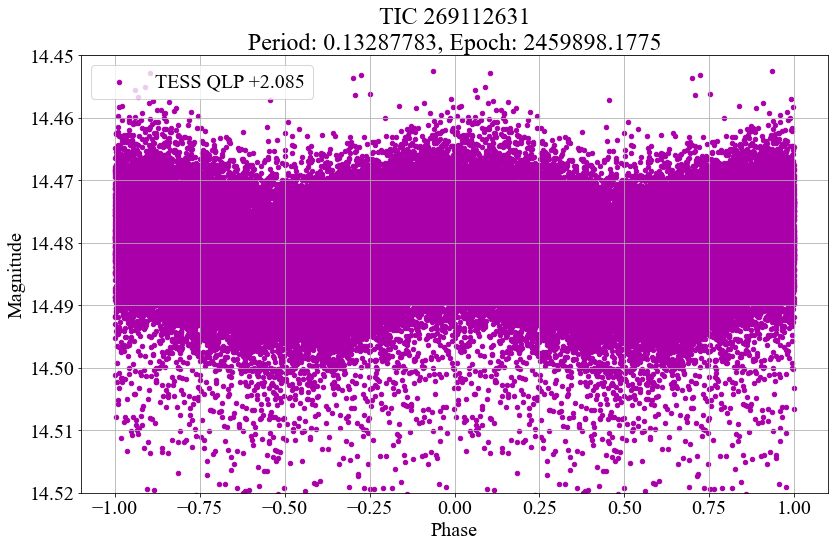}
\caption{Phase-folded light curve of TIC 269112631 (pulsational variability).}
\label{figPMAK_V150puls}
\end{figure}

We then performed periodogram analysis of the light curve excluding the minima. The periodogram shows a strong peak at a frequency of 7.5257~$d^{-1}$; see Fig.~\ref{figPMAK_V150periodogram}. The refined period corresponding to this frequency is 0.13287783(9)~d. The corresponding phase plot is shown in Fig.~\ref{figPMAK_V150puls}.

% png format

% png format

The star has two neighbors, TIC 269112615, $22''$ away, and TIC 269112612, $23''$ away. They are close enough to blend with the target (the TESS pixel scale is $21''$ per pixel; \citealp{ricker2014}). To ensure that the observed variability originates from the target rather than from neighboring stars, we performed direct aperture photometry of the TESS image using one-pixel apertures with the help of Lightkurve for Sector 17.

\begin{figure}[htbp]
\centering
\includegraphics[width=15cm]{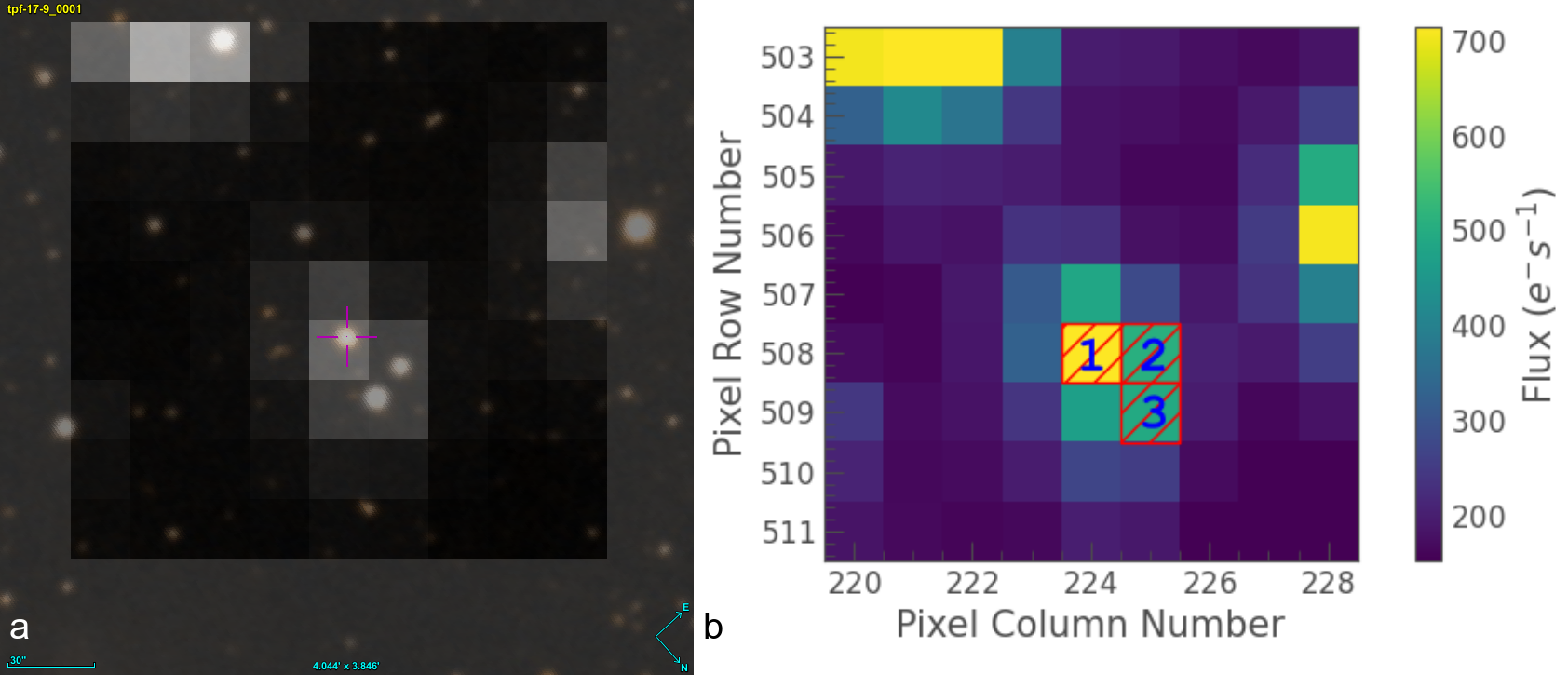}
\caption{(a) Cutout of a TESS full-frame image for Sector 17, with the star field from Aladin overlaid. The target star, TIC 269112631, is marked with crosshairs. (b) Three one-pixel apertures used for test photometry.}
\label{figPMAK_V150TESSimage}
\end{figure}

\begin{figure}[htbp]
\centering
\includegraphics[width=12cm]{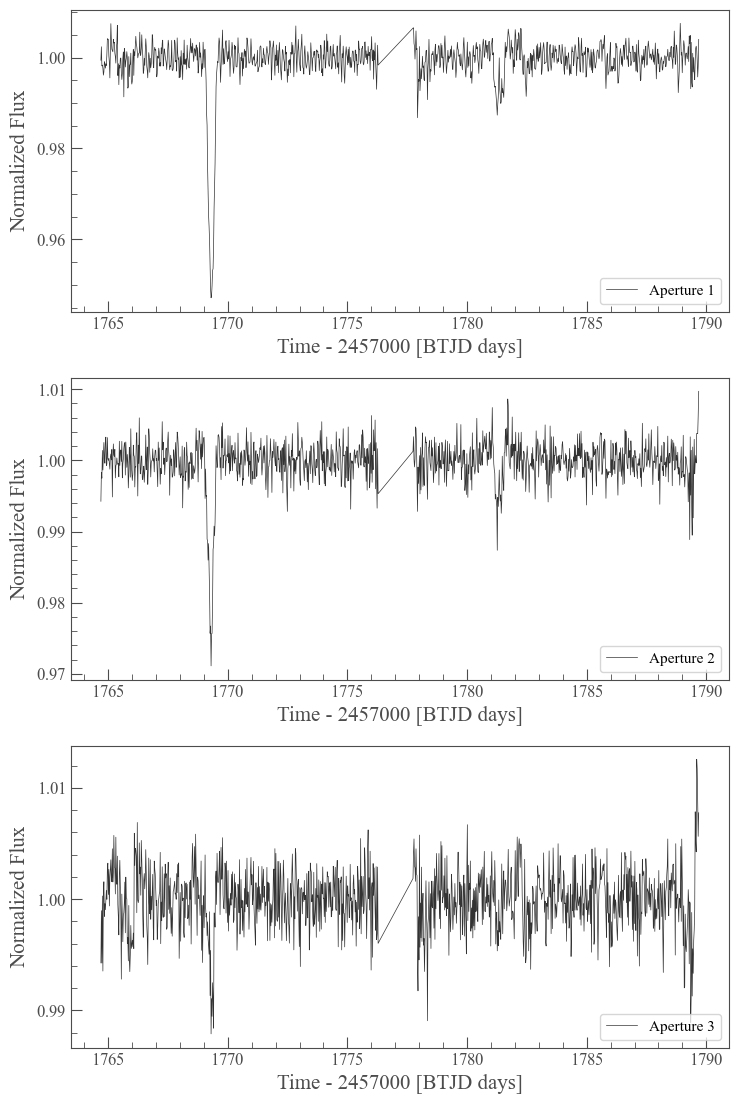}
\caption{Flattened and normalized light curves obtained from Lightkurve photometric analysis of TIC 269112631.}
\label{figPMAK_V150lk}
\end{figure}

Fig.~\ref{figPMAK_V150TESSimage}a shows a cutout of a TESS full-frame image for Sector 17, with the star field from Aladin \citep{bonnarel2000} overlaid. Fig.~\ref{figPMAK_V150TESSimage}b shows three one-pixel apertures we used for the test. The photometry results obtained with these apertures are shown in Fig.~\ref{figPMAK_V150lk}. The light curves were flattened and normalized using the standard Lightkurve \textit{flatten} procedure. It can be seen that the most prominent eclipses are detected for aperture 1, which is positioned directly on the target star. The same applies to the pulsations: for aperture 1, the peak in the periodogram at the frequency corresponding to a pulsation period of 0.13287783 d is strong; it is also distinguishable but noticeably weaker for aperture 2, and is not detectable for aperture 3. This demonstrates that both the eclipses and the pulsations originate from the target star.

The star exhibits large color indices: $B-V = 1.58$ \citep{henden2015}, $J-K = 0.81$ \citep{cutri2003}, and $BP-RP = 2.27$ \citep{gaia2022a}. At the same time, the extinction coefficient $A_V$ for this star is 5.2 \citep{khalatyan2024}, which can be explained by its location in a dust-rich region near the Bubble Nebula. Adopting the extinction law of \citep{cardelli1989}, with updates by \citep{odonnell1984}, a total-to-selective extinction ratio of $R_V = 3.1$ \citep{fitzpatrick1999}, and using extinction coefficients for Gaia passbands from \citep{wang2019}, we estimated the intrinsic color indices: $(B-V)_0 \approx -0.10$, $(J-K)_0 \approx -0.05$, and $(BP-RP)_0 \approx 0.12$. These values are compatible with a late B or early A spectral type.

\begin{figure}[htbp]
\centering
\includegraphics[width=12cm]{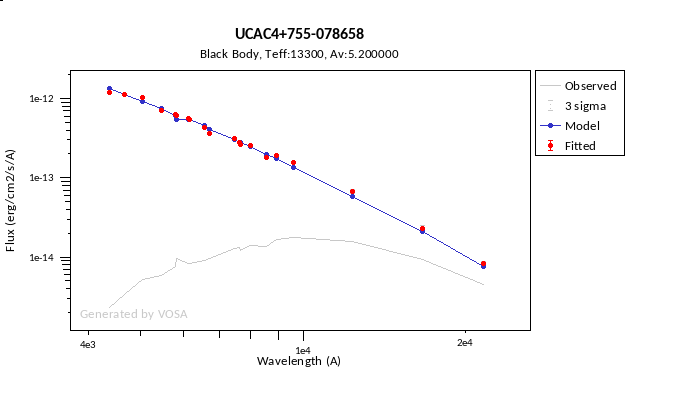}
\caption{Spectral energy distribution (SED) of TIC 269112631. The gray line shows observed data, not corrected for extinction.}
\label{figPMAK_V150SED}
\end{figure}

We also calculated the spectral energy distribution (SED) using the online tool VOSA \MyLink{https://svo2.cab.inta-csic.es/theory/vosa/} \citep{bayo2008}. The SED was fitted using a black-body model approximation, and the result is shown in Fig.~\ref{figPMAK_V150SED}. The extinction coefficient $A_V = 5.2$ (adopted from \citealp{khalatyan2024}, as specified above) was used. The effective temperature obtained is $T_{\mathrm{eff}} = 13300$ K, which corresponds to a late B-type star.

Using the Gaia DR3 parallax of 0.333 milliarcseconds \citep{gaia2022a}, we estimate an approximate distance of 3 kpc. The apparent magnitude derived from the Gaia DR3 data is $V = 14.48$. Taking into account the extinction $A_V = 5.2$, we obtain an absolute magnitude of $M_V \approx -3.1$, which is typical for an early B-type star.

There is a discrepancy between the spectral type inferred from the color-based effective temperature and SED, and that implied by the luminosity. This discrepancy is most likely because the object is a binary system, where the color indices and SED are affected by the combined light of both components. In this case, the derived temperature is biased toward lower values due to the contribution of the cooler companion. Overall, the system is consistent with a B-type star.

Given the observed short-period ($\approx 0.133 d$), stable variability, amplitude of about 0.01 TESS, the star is a strong candidate for a $\beta$ Cephei-type pulsator. The pulsation period and amplitude are consistent with this classification (see \\ \MyLink{https://vsx.aavso.org/index.php?view=about.vartypes\#BCEP}).

% png format

% png format

% png format

\subsection{TIC 269809093}
\begin{figure}[htbp]
\centering
\includegraphics[width=15cm]{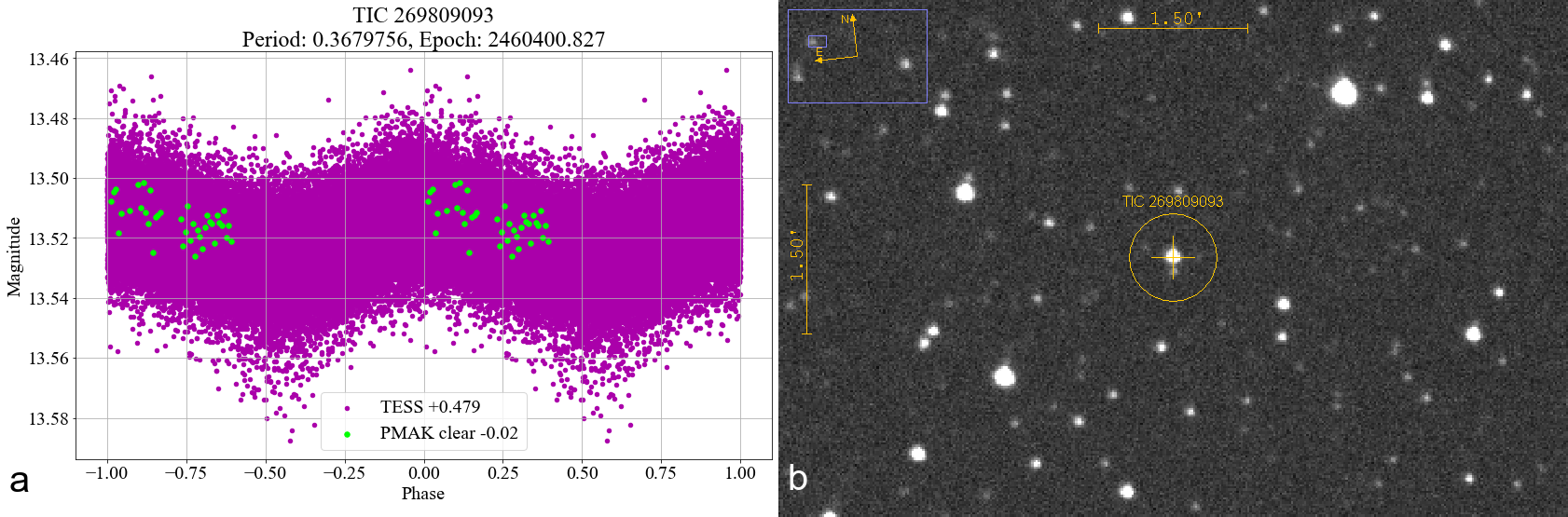}
\caption{(a) Phase-folded light curve of TIC 269809093; (b) the corresponding field.}
\label{figPMAK_V151}
\end{figure}

This star is registered in the VSX database as PMAK V151. In Fig.~\ref{figPMAK_V151}a, the phase plot is shown. Our data are labeled as PMAK. The corresponding star field is shown in Fig.~\ref{figPMAK_V151}b. The variability type classification (slowly pulsating B star, SPB) is based on the star’s spectral type from \citep{creevey2023}, as well as its period and variability range.

TESS QLP light curves from Sectors 57, 58, 77, 78, 84, and 85 were used to calculate the period and initial epoch. The value of the period was refined with VS-fit using a second-degree trigonometric polynomial. The light curves were aligned to the median magnitude derived from Gaia DR3 data.

\begin{figure}[htbp]
\centering
\includegraphics[width=12cm]{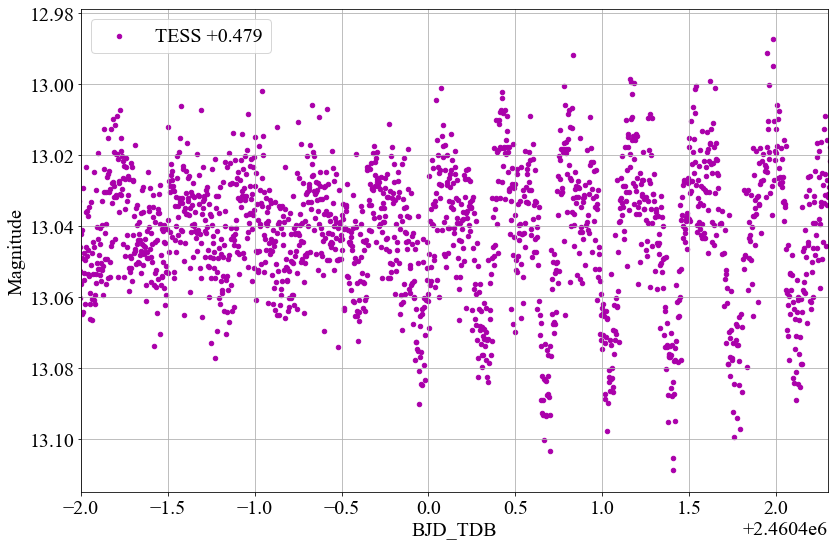}
\caption{Part of the light curve of TIC 269809093 in Sector 77.}
\label{figPMAK_V151lc}
\end{figure}

The light curve has a complex shape, as can be seen in Fig.~\ref{figPMAK_V151lc}. As a result, the period and initial epoch values listed in Tab.~\ref{tableStars} may vary.

% png format

% png format

\subsection{TIC 269112450}
\begin{figure}[htbp]
\centering
\includegraphics[width=15cm]{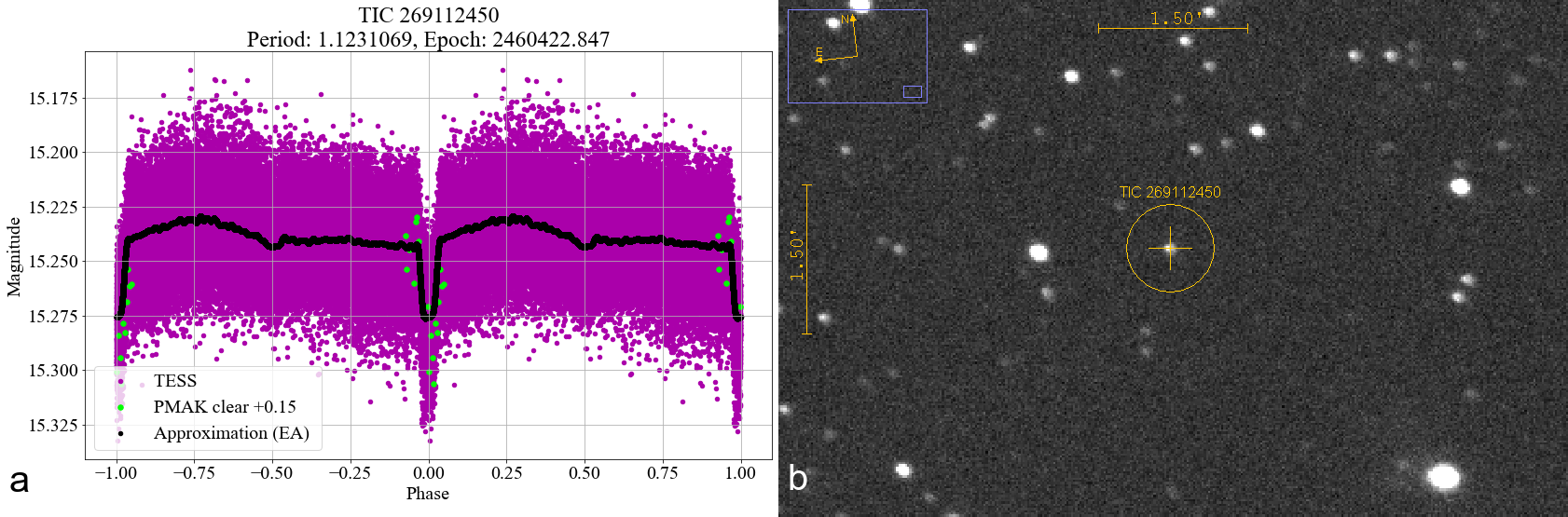}
\caption{(a) Phase-folded light curve of TIC 269112450; (b) the corresponding field.}
\label{figPMAK_V152}
\end{figure}

This star is registered in the VSX database as PMAK V152. It is an eclipsing variable with additional rotational variability, consistent with the BY Draconis-type with Algol-type eclipses (EA+BY). Fig.~\ref{figPMAK_V152}a shows the phase plot, with our data labeled as PMAK. The light curves were aligned using the mean V magnitude from Gaia DR3. The corresponding star field is shown in Fig.~\ref{figPMAK_V152}b.

No pre-calculated light curves were available; therefore, we performed photometry on TESS full-frame images using Lightkurve. Data from Sectors 17, 24, 57, 58, 77, 78, 84, and 85 were used. We removed long-term trends and aligned the light curves across sectors using LCV, and then estimated the periods of the EA and BY variability from the phase-folded light curve by eye.

The light curve was then approximated by a 50th-degree trigonometric polynomial for the EA variability, combined with a second-degree trigonometric polynomial for the BY variability, with the periods refined using VS-fit. The component of the model corresponding to the EA variability is shown in Fig.~\ref{figPMAK_V152}a. In addition to the primary eclipse, a weak secondary minimum is also captured by the model, though it is not clearly apparent in the raw data.

\begin{figure}[htbp]
\centering
\includegraphics[width=12cm]{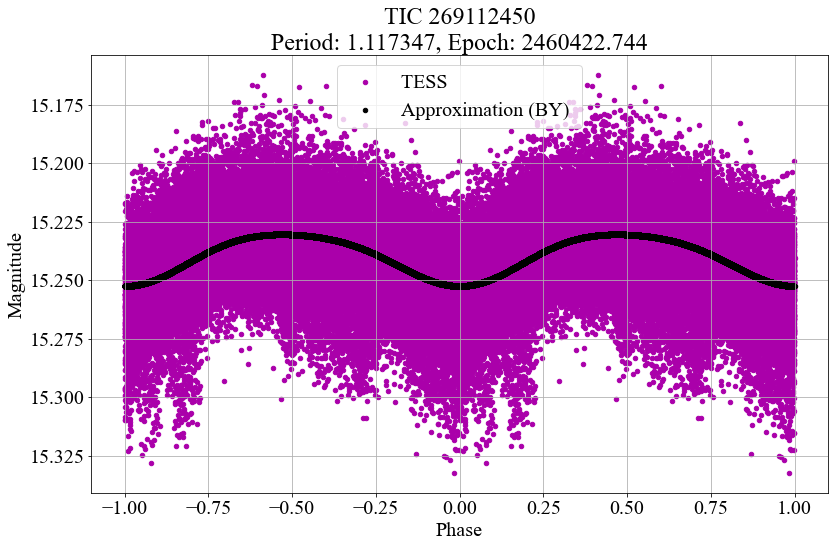}
\caption{Phase-folded light curve of TIC 269112450 showing BY-type variability.}
\label{figPMAK_V152by}
\end{figure}

The phase plot for the BY variability, together with the corresponding model component, is shown in Fig.~\ref{figPMAK_V152by}.

\begin{figure}[htbp]
\centering
\includegraphics[width=15cm]{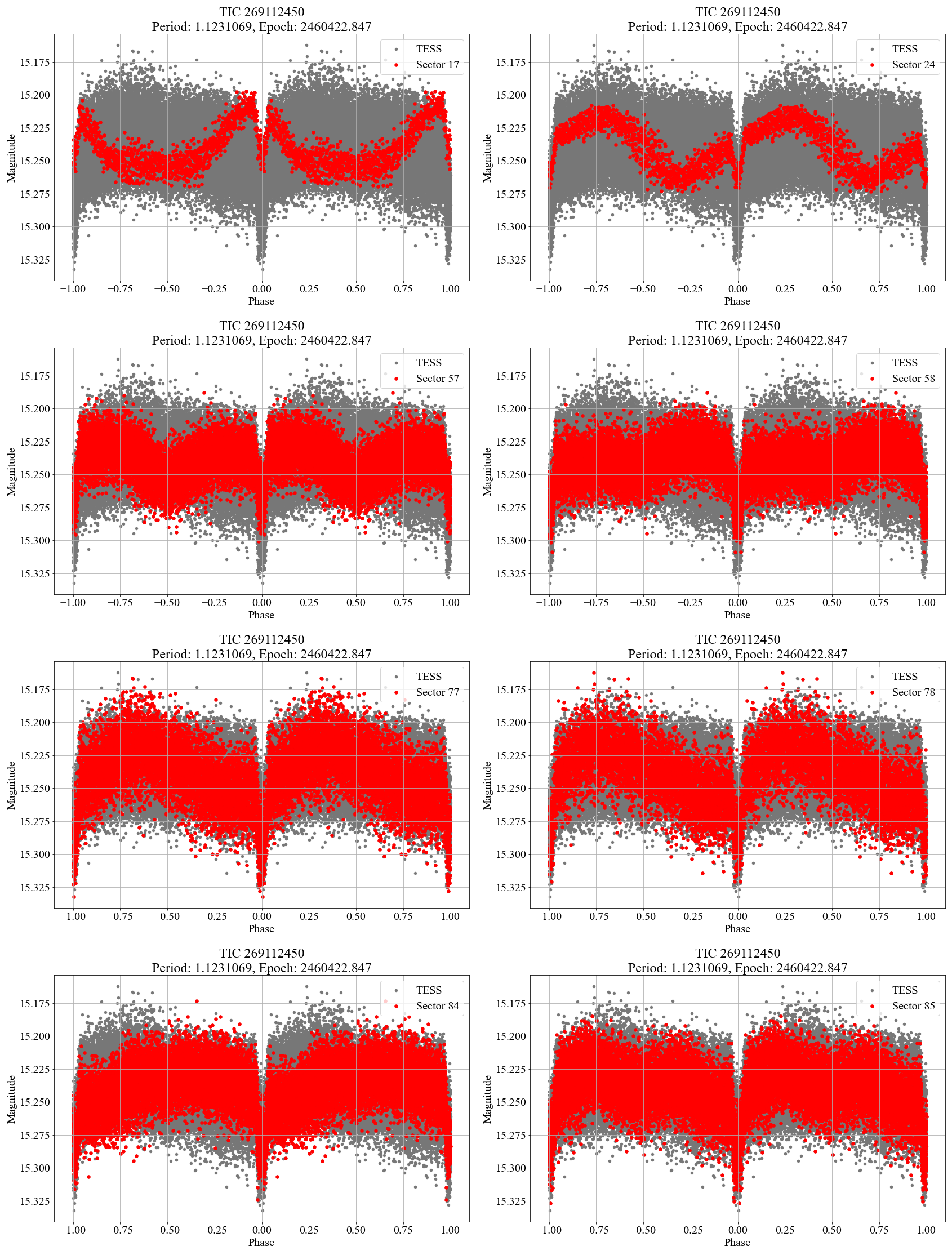}
\caption{Light curves of TIC 269112450 in different TESS sectors.}
\label{figPMAK_V152lc}
\end{figure}

Due to the combined variability, the light curve has a complex shape and evolves over time. Phase-folded light curves for individual TESS sectors are shown in Fig.~\ref{figPMAK_V152lc}. Sectors 17 and 24 are less noisy because of their longer effective exposure times (1426 s), compared to 158 s in the remaining sectors.

% png format

% png format

% png format

\section{Conclusions}\label{secconc}

Using ground-based observations with a small-aperture telescope, we discovered six new variable stars that were missed by automated survey pipelines.

Further classification and characterization of these variables were performed using photometric data from the Transiting Exoplanet Survey Satellite (TESS). The analysis was carried out using our open-source software tools, LCV and VS-fit. These tools proved to be efficient for periodogram analysis and light-curve modeling.

The star TIC 269112631 may be an interesting astrophysical target because it is an eclipsing binary that exhibits pulsations. Such systems are valuable because they allow simultaneous constraints to be placed on fundamental stellar parameters and internal structure, making this star a promising target for further study.

Another eclipsing star described in this article, TIC 269112450, is an interesting example of a totally eclipsing BY Draconis-type binary in which starspot-induced rotational variability is comparable to, or even exceeds, the eclipsing variability. This system could be a good candidate for further research into highly active binary systems.

\setcounter{secnumdepth}{0}
\MyAcknowledgements{
This paper includes data collected by the Transiting Exoplanet Survey Satellite (TESS) mission, funded by NASA. This research has made use of data products from the Quick-Look Pipeline (QLP) and the TESS Science Processing Operations Center (SPOC), obtained from the Mikulski Archive for Space Telescopes (MAST). Part of the analysis was carried out using the Lightkurve package.
This publication makes use of VOSA, developed under the Spanish Virtual Observatory (\MyLink{https://svo.cab.inta-csic.es}) project funded by MCIN/AEI/10.13039/501100011033/ through grant PID2020-112949GB-I00.
VOSA has been partially updated by using funding from the European Union's Horizon 2020 Research and Innovation Programme, under Grant Agreement nº 776403 (EXOPLANETS-A) 
The author would like to express sincere gratitude to Prof. Ivan L. Andronov for his invaluable advice, continuous support, and inspiration, as well as for his careful comments and corrections that greatly improved this work. The author is especially grateful to Sebastian Otero, administrator of the VSX database, for his kind support and guidance during the registration of the new variable stars.} \\

%\newpage

\end{document}